\documentclass[twocolumn,amssymb,fleqn]{revtex4} %fleqn: make all eqns left alighed  ,showpacs
\usepackage{epsfig,amssymb,amsmath,graphicx,subfigure,hyperref  }  % pdfpages and graphicx are added   hyperref(set hyperlink for content)
\usepackage{color}

\usepackage{multirow}
\usepackage{tabularx}

\begin{document}

\title{Harmonic field in knotted space}
\author{Xiuqing Duan and Zhenwei Yao}
\email{zyao@sjtu.edu.cn}
\affiliation{School of Physics and Astronomy, and Institute of Natural
Sciences, Shanghai Jiao Tong University, Shanghai 200240, China}
\begin{abstract}
Knotted fields enrich a variety of physical phenomena, ranging from fluid flows, electromagnetic
fields, to textures of ordered media. Maxwell's electrostatic equations, whose vacuum
solution is mathematically known as a harmonic field, provide an ideal setting to explore
the role of domain topology in determining physical fields in confined space.
In this work, we show the uniqueness of a harmonic field in knotted tubes, and reduce the
construction of a harmonic field to a Neumann boundary value problem. By analyzing the harmonic
field in typical
knotted tubes, we identify the torsion driven transition from bipolar to vortex
patterns. We also analogously extend our discussion to the organization of
liquid crystal textures in knotted tubes.  These results further our understanding about
the general role of topology in shaping a physical field in confined space, and may find
applications in the control of physical fields by manipulation of surface topology. 
\end{abstract}

\maketitle

Understanding physical fields in confined space is a common theme in a host
of scientific problems, ranging from the classical examples in
hydrodynamics~\cite{kleckner2013creation,scheeler2014helicity,scheeler2017complete} and
electrodynamics~\cite{kedia2013tying}, to the fabrication of geometrically
confined liquid crystals for various applications~\cite{bisoyi2011liquid,
alexander2012colloquium, umadevi2013large,urbanski2017liquid}. The topology of the domain can
critically determine the configuration of a physical field~\cite{klauder1957question,
enciso2015existence}. Of special interest is the structure of a physical field filling
knotted tubes. Knotted field configurations, previously known in Lord Kelvin's
theoretical proposal of the vertex atom hypothesis inspired by the work of
Helmholtz~\cite{thomson1867ii}, have been experimentally accessible in diverse physical
and chemical systems~\cite{horner2016knot}, including vortex loops in
superconductors~\cite{samokhvalov1996vortex, tes̆anovic1999extreme}, defect loops in
liquid
crystals~\cite{smalyukh2010three,tkalec2011reconfigurable,vcopar2011nematic,vcopar2013visualisation,
martinez2014mutually,vcopar2014topology,vcopar2015knot,ackerman2017diversity}, toroidal nematic
textures~\cite{Pairam2013,
chen2013generating,koning2014saddle,koning2014saddle,ackerman2015self, urbanski2017liquid}
and knotted beams of light~\cite{irvine2008linked,dennis2010isolated,irvine2010linked}.
Note that in these systems knotted fields mostly occur either in vacuum space or in the free
space of viscous fluids and nematic liquid crystals. Past studies have
shown that Maxwell's equations, despite their linearity, can admit topologically nontrivial
knotted solutions in free space~\cite{bateman1915mathematical, penrose1967twistor,
ranada1989topological,ranada1992topological,
irvine2008linked,irvine2010linked,hoyos2015new,kedia2016weaving}. These results suggest that the system of
Maxwell's equations provides an ideal setting to address the fundamental question of how
topology shapes behaviors of physical fields in confined space.

The goal of this work is to construct vacuum solutions to Maxwell's electrostatic
equations $\textrm{div}\ \mathbf{E}=0$ and $\textrm{curl}\ \mathbf{E}=0$ with tangential
boundary condition in topologically nontrivial domains of knotted tubes. Such a solution
is mathematically known as the harmonic field~\cite{enciso2015existence}.  The existence of
nontrivial smooth harmonic field in confined space depends on the topology of the
domain~\cite{enciso2015existence}. As an example, a smooth field is forbidden in spherical closed
space; a singularity in the field is inevitable. However, toroidal space can admit a nontrivial vacuum
solution~\cite{klauder1957question}.

In this work, we first show the uniqueness of harmonic field as dictated by the topology
of knotted tube. By introducing an irrotational field, we reduce the construction of
harmonic field to solving a Neumann boundary value problem.  The harmonic field in the
circular tube (a standard torus) is analytically derived. For a torsion free,
elliptic torus with spatially varying curvature along its core loop, we find that the
harmonic vector field becomes tilted, and its projection to the cross-section of the tube
exhibits a bipolar configuration. By further examining the projected harmonic field in
the cross-sections of trefoil and cinquefoil torus knots, we identify the torsion driven
transition from the bipolar to the vortex configurations. These results reflect the
general feature of a knotted harmonic field that is beyond the specific Maxwell's
equation system. As an example, we finally extend our discussion to the organization of
nematic textures confined in knotted tubes.

\begin{figure}[b]
  \includegraphics[width = 3in, bb=12 87 600 480]{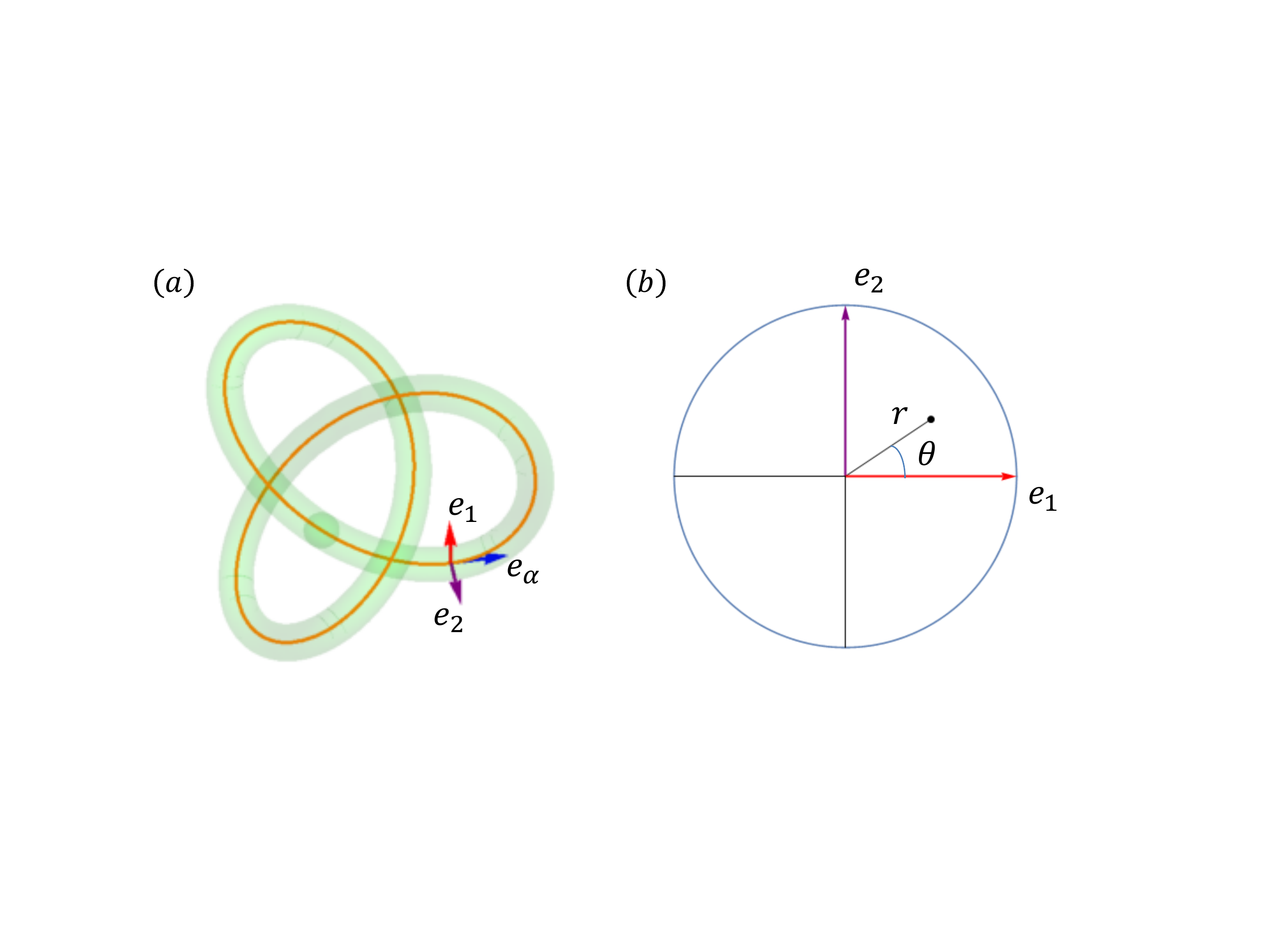}
  \caption{Illustration of a knotted tube constructed out of the red core loop. (a) The
  tangent vector $\mathbf{e_{\alpha}}$, normal vector $\mathbf{e_1}$, and binormal vector
  $\mathbf{e_2}$ form a moving frame of reference named the Frenet--Serret frame along the
  core loop. 
  (b) The pair of the normal and binormal vectors define a two-dimensional unit
  disk.  } 
\end{figure}

  \begin{figure*}[t] 
  \includegraphics[width = 2\columnwidth]{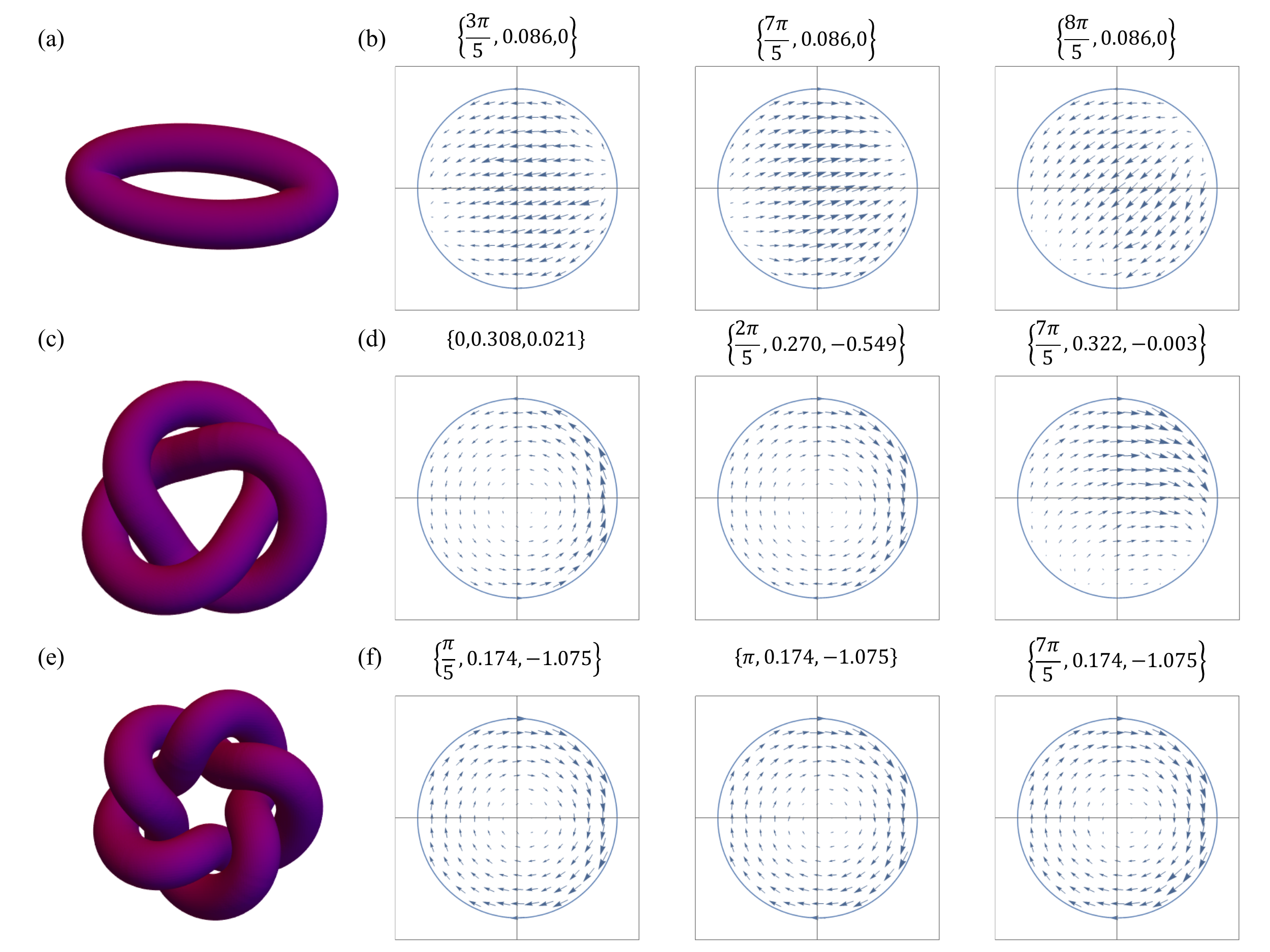}
  \caption{Visualization of the harmonic field projected to the cross-section of
  typical knotted tubes.  (a)-(b) Solid ellipse torus: $(4.6
  \textrm{cos} t,1.6 \textrm{sin} t,0)$. (c)-(d) Trefoil knot:
  $(10/9)[(3+\textrm{cos} (3t)) \textrm{cos} (2t),(3+\textrm{cos} (3t))
  \textrm{sin} (2t), \textrm{sin} (3t)]$. (e)-(f) Cinquefoil torus knot:
  $((3+1.1 \textrm{cos} (5t)) \textrm{cos} (2t),(3+1.1 \textrm{cos} (5t))
  \textrm{sin} (2t), 2 \textrm{sin} (5t))$.  The triple numbers in the curly
  brackets are the values for $t$, the curvature $\kappa(t)$, and the torsion
  $\tau(t)$, respectively. $t \in [0,2\pi)$. The thickness of all these knots is
  $\epsilon=1$. } 
\end{figure*}

{\bf{Frenet--Serret parametrization of knotted tube} } We construct the domain of filled knotted tube
from a closed loop $\gamma$; it is named the core loop of the tube. A solid torus is
a special case of a filled knotted tube. The arc-length parametrization of the loop
$\gamma$ is $\gamma:\mathbb{S}_L \rightarrow \mathbb{R}^3$, where $\mathbb{S}_{L}
=[0,L]$, and $L$ is the length of the loop. The domain inside the tube is denoted as
$T_\epsilon (\gamma)$.  $T_\epsilon (\gamma) = \{x \in \mathbb{R}^3 : \textrm{dist}(x,
\gamma) < \epsilon \}$, where $\textrm{dist}(x, \gamma)$ is the distance between the
point $x$ and the loop $\gamma$, and $\epsilon$ is the thickness of the tube. $\epsilon$
is positive and sufficiently small to avoid self-intersect of the knotted tube.

To represent any point within the tube, we introduce the triple of
$\{\dot{\mathbf{\gamma}}(\alpha), \mathbf{e_1}(\alpha), \mathbf{e_2(\alpha)}\}$ pointwise
along the core loop, which are the tangent, normal and binormal unit vectors, as shown in
Fig.1(a). These unit vectors form a moving frame of reference 
named the Frenet--Serret
frame along the core loop~\cite{Nakahara,panoramic}.
The arc-length parameter $\alpha \in [0, L)$.
The pair of the normal and binormal vectors
define a two-dimensional unit disk denoted by $D^2$ [see Fig.1(b)]. Any point in
$T_\epsilon$ can therefore be represented by the coordinates $(\alpha, y_1,
y_2)\in \mathbb{S}_L \times D^2$ via the
diffeomorphism $(\alpha, y_1, y_2) \longmapsto \mathbf{\gamma} (\alpha) + \epsilon y_1
\mathbf{e_1}(\alpha) +\epsilon y_2 \mathbf{e_2}(\alpha)$. $y_1, y_2 \in [0, 1]$. The Euclidean
metric in the tube is 
\begin{eqnarray}
  ds^2 = A d\alpha^2+2 \epsilon^2 \tau (y_2 dy_1 -
y_1 dy_2) d\alpha + \epsilon^2 (dy_1^2 + dy_2^2),\nonumber
\end{eqnarray}
  where $A = (1-\epsilon \kappa y_1)^2 + (\epsilon \tau)^2 (y_1^2+y_2^2)$, $\kappa
  \equiv \kappa(\alpha)$ and $\tau \equiv \tau(\alpha)$ are the curvature and
  torsion of the core loop $\gamma$. The volume element $dV = \epsilon^2 B
  d\alpha dy_1dy_2$, where $B = 1-\epsilon \kappa y_1$.

{\bf{Construction of harmonic field} } All the
harmonic fields, denoted as $\mathbf{h}$ in the filled knotted tube $T_{\epsilon}$
constitute a vector space~\cite{enciso2015existence}: $\label{harmonic field}
\mathcal{H}(T_{\epsilon}) = \{\mathbf{h} \in C^{\infty}(T_{\epsilon}, \mathbb{R}^3) :
\textrm{div} \ \mathbf{h} = 0, \textrm{curl} \ \mathbf{h} = 0 \ \textrm{and} \
\mathbf{h} \cdot \boldsymbol{\nu} = 0 \}$, where $C^{\infty}(T_{\epsilon}, \mathbb{R}^3)$ denotes an
infinitely differentiable functional space defined in $T_{\epsilon}$, and $\boldsymbol{\nu}$
is the unit outward normal vector on the surface of $T_{\epsilon}$. 

We first show the uniqueness of the harmonic field (up to a multiplicative constant) as a
fundamental consequence of the topological structure of the space $T_{\epsilon}$. The
space $\mathcal{H}(T_{\epsilon})$ is isomorphic to the first cohomology group
$H^1(T_{\epsilon})$ of the filled knotted tube $T_{\epsilon}$, which is associated with
the only non-contractible loop along the curve $\gamma$~\cite{Nakahara,
enciso2015existence}. Furthermore, $H^1(T_{\epsilon}) \cong H^1(S^1 \times D^2) \cong
H^1(S^1) \cong \mathbb{Z}$, where $S^1$ is a topological circle and $D^2$ is a
topological disk~\cite{Nakahara}. Therefore, the space $\mathcal{H}(T_{\epsilon})$ is
one-dimensional in the sense that if $\mathbf{h}$ is a harmonic vector field, then all
the other harmonic vector fields are linear with $\mathbf{h}$~\cite{Nakahara,
enciso2015existence}. Here, we emphasize that the topology of the filled knotted tube
$T_{\epsilon}$ determines the existence and uniqueness of the harmonic field, which lays
the foundation for the following construction of the harmonic field.

To construct the harmonic field in $T_{\epsilon}$, we consider the vector field
$\mathbf{h_0} = B^{-2}(\partial_{\alpha}+ \tau \partial_{\theta})$, where $B = 1-\epsilon
\kappa y_1$, and $\theta$ is the polar angle as shown in Fig.1(b). One can show that
$\mathbf{h_0}$ is irrotational and satisfies the tangential boundary condition (the
proof is given in Appendix A).  Based on $\mathbf{h_0}$, we construct a harmonic vector
field by the Hodge decomposition~\cite{Nakahara, enciso2015existence}: \begin{eqnarray}
\mathbf{h} = \mathbf{h_0} + \nabla \psi,\label{h} \end{eqnarray}
The divergence-free condition requires that the scalar function $\psi$ in
Eq.(\ref{h}) must satisfy
\begin{eqnarray}
  \Delta \psi = \varrho \label{N1}
  \end{eqnarray}
in $T_{\epsilon}$ with the boundary condition $\partial_{\boldsymbol{\nu}}
\psi|_{\partial T_{\epsilon}} = 0$.
$\varrho = - \textrm{div}\ \mathbf{h_0} =  \varepsilon B^{-3} r (\tau \kappa \textrm{sin} \theta - \kappa'
\textrm{cos} \theta)$, satisfying $\int
\varrho dV =0$. The solution $\psi$ is unique up to a constant. Therefore, the
search for the harmonic vector field in $T_{\epsilon}$ is nicely reduced to the
Neumann boundary value problem in Eq.(\ref{N1}).

The harmonic vector field can be derived analytically when the core loop
$\gamma$ is a circle, i.e., the tube is a standard solid torus. Since $\tau=0$ and
$\kappa$ is a constant, the source term $\varrho$ in Eq.(\ref{N1}) vanishes.
Multiplying Eq.(\ref{N1}) by $\psi$, we have \begin{align} \int_{T_{\epsilon}}
\psi \triangle \psi dV =& \int_{\partial T_{\epsilon}} \psi
\partial_{\boldsymbol{\nu}} \psi dS - \int_{T_{\epsilon}} |\nabla \psi|^2 dV
\nonumber \\ =& -\int_{T_{\epsilon}} |\nabla \psi|^2 dV = 0.\nonumber \end{align}
Requiring $\nabla \psi=0$ leads to the expression for the harmonic field:
\begin{align} \mathbf{h}=B^{-1} \mathbf{e_{\alpha}} = \frac{1}{1-\epsilon \kappa y_1}
  \mathbf{e_{\alpha}},  \label{mag} \end{align} where $\mathbf{e_{\alpha}}$ is the unit tangent
  vector to the core loop $\gamma$ [see Fig.1].  We recognize that Eq.~(\ref{mag}) has the
  same functional form as that of the magnetic field generated by electric
  current in a straight wire.

For an elliptic core loop with spatially varying curvature $\kappa(\alpha)$, we
numerically solve Eq.(\ref{N1}), and find that the harmonic vector field becomes tilted with a nonzero
transverse component $\mathbf{h}_{\perp}$ lying over the cross-section of the
tube, as shown in Fig.2(b). In contrast, $\mathbf{h}_{\perp}=0$ for the
case of a standard solid torus.  The triple numbers in the
curly bracket in Fig.2 are the values for $t$, the curvature $\kappa(t)$, and the torsion
$\tau(t)$, respectively. $t$ is a parametrization of the core loop. $t \in [0,
2\pi)$, corresponding to $\alpha \in[0, L)$.  From Fig.2(b), we see that the tilted harmonic
vector field $\mathbf{h}_{\perp}$ exhibits a bipolar configuration, and the
entire field configuration rotates along the core loop. The strength of the
$\mathbf{h}_{\perp}$ field approaching the diametric poles becomes
vanishingly small, which is consistent with the tangential boundary condition.

To examine the effect of torsion on the field configuration, we further consider
knotted tubes constructed out of core loops with torsion. In Fig.2(c) and
2(e), we show the trefoil and cinquefoil torus knots with crossing number
three and five, respectively~\cite{adams2004knot}. Their
names are from the three-leaf clover plant and the five-petaled flowers of
plants in the genus Potentilla. The trefoil knot is the simplest example of a
nontrivial knot. From Fig.2(d) and 2(f), we see that the configuration of the
$\mathbf{h}_{\perp}$ field becomes azimuthal, which is distinct from that in
Fig.2(b) for the case of a torsion free core loop.

We further notice that, at the location of small torsion on the core loop, as shown in
the last plot of Fig.2(d), the $\mathbf{h}_{\perp}$ field is in an intermediate state
between the bipolar and vortex configurations. All these observations substantiate the
physical scenario of torsion driven transition from the bipolar to the vortex
configurations.  In all the torsion-driven vortex structures in Figs.2(d) and 2(f),  the
field strength near the center becomes zero to avoid singularity. It is of interest to
note that the transformation of the $\mathbf{h}_{\perp}$ field from the bipolar to the
vortex structure resembles the merge of a pair of +1/2 defects into a single +1 defect in
nematic textures over spherical disks~\cite{duan2017curvature}.

{\bf{Nematic texture in knotted tube} } Now, we extend our discussion to the system of
nematic liquid crystal (LC) confined in a knotted tube. Self-assembly of LC in various
confined environments especially within the cylindrical polymer sheath represents a new
trend in LC researches for the promising applications in the new generation of wearable
technology devices~\cite{urbanski2017liquid}. The experimentally accessible
system of a LC filled knotted tube is an ideal model to address the inquiry into the
organization of matter by the topology of the domain. In the following, we discuss how our preceding
discussions on harmonic field yield insights into this question. We consider
a planar boundary condition where LC molecules at the boundary lie in the
tangent plane~\cite{kleman2007soft}.

In the continuum limit, the orientations of LC molecules are characterized by a
director field $\mathbf{n}(x)$. $\mathbf{n}$ is a unit vector and
$\mathbf{n}\equiv -\mathbf{n}$ due to the apolar nature of LC molecules. 
According to the Frank free energy model for nematics, the free energy cost
associated with the deformation of the director field from the uniform state is
~\cite{kleman2007soft}
\begin{eqnarray} 
    F[\mathbf{n(x)}] =  \int f dV - K_{24}\int d \mathbf{S} \cdot \mathbf{g_{24}},
    \label{F}
\end{eqnarray} 
where $f = \frac{1}{2} K_1 (\nabla \cdot \mathbf{n})^2 + \frac{1}{2}  K_2 (\mathbf{n}
\cdot \nabla \times \mathbf{n})^2 +   \frac{1}{2} K_3 (\mathbf{n} \times \nabla \times
\mathbf{n})^2$. $K_1, K_2$, and $K_3$  are the splay, twist, and bending moduli,
respectively. In the surface term, $K_{24}$ is the saddle-splay
modulus, and $d\mathbf{S} =\boldsymbol{\nu} dS$ is the area element,
where $\boldsymbol{\nu}$ is
the outward unit normal vector on the surface. $\mathbf{g_{24}} = \mathbf{n}\nabla \cdot
\mathbf{n}+\mathbf{n}\times \nabla\times \mathbf{n}$.

Both the volume and surface terms in the expression for the Frank free energy $F$ in
Eq.(\ref{F}) vanish when $\mathbf{n}$
is a harmonic field whose divergence and curl are zero. However, our preceding
discussion shows that the expression for the harmonic field takes the form of
$\mathbf{h}=B^{-2}(\partial_{\alpha}+ \tau \partial_{\theta})+\nabla \psi$, and it is not
a unit vector, which is in conflict with the condition of $|\mathbf{n}|=1$. 
Consequently, the Frank free energy of the ground state nematic texture
in a knotted tube must be nonzero, as dictated by the topology of the
domain. It is of interest to study the minimization of the Frank free energy via the
interplay of the director field and the geometry of the knotted tube. As an analytically tractable
case, we work in the constraint of an untilted director field, and derive that a torsion-free tube tends to take a circular shape to minimize the free energy.  The details are
presented in Appendix B.

In summary, we study the problem of the harmonic field confined in knotted space that is
inspired by solving for the vacuum solution to Maxwell's electrostatic equations. We show
that the topology of a knotted tube determines the existence and uniqueness of the harmonic field, and reduce the
construction of a harmonic field to a Neumann boundary value problem. From the solved harmonic field in
typical knotted tubes, we identify the torsion driven transition from bipolar to vortex
patterns. We also analogously extend our discussion to the organization of liquid crystal
textures in knotted tubes. These results further our understanding about how topology
shapes behaviors of physical fields in confined space, and may find
applications in the control of physical fields by the manipulation of surface topology.

\section*{Appendix A: Proof of the irrotational nature of $\mathbf{h_0}$}

The coordinate-independent expression for the curl operator is 
$\textrm{curl}~\mathbf{n}=(\star d
n^{\flat})^{\sharp}$, where $\star$ is an
operator called Hodge dual, and $\scriptstyle\flat$ and $\scriptstyle\sharp$ are the
musical isomorphisms~\cite{Nakahara,panoramic,multivariable}. We calculate curl $\mathbf{h_0}$ step by step. 
$\mathbf{h_0}^{\flat}= AB^{-2}d\alpha +(-\epsilon^{2} \tau r^2)B^{-2}d\theta + (-\epsilon^2 \tau r^2)B^{-2} \tau d\alpha+\epsilon^2 r^2 \tau B^{-2}d\theta = d\alpha $, where the last equality is by inserting $A$ and $B$, whose expressions are
given in the main text. So we have
$d h_0^{\flat}=dd\alpha=0$. Consequently, curl~$\mathbf{h_0}=(\star d h_0^{\flat})^{\sharp}=0$.

\section*{Appendix B: Optimal geometry of torsion-free, nematics-filled tube}

In this section, we will show that a torsion-free tube filled with nematics tends to take a
circular shape to minimize the Frank free energy. We first represent the director
$\mathbf{n}$ by $\mathbf{n}=n_1(\alpha,r,\theta)\mathbf{e_{\alpha}}+n_2(\alpha,r,\theta)
\mathbf{e_{r}}+n_3(\alpha,r,\theta)\mathbf{e_{\theta}}$, 
where $\mathbf{e_{\alpha}}=\frac{1}{\sqrt{A}}\partial_{\alpha}$,
$\mathbf{e_{r}}=\frac{1}{\epsilon}\partial_{r}$, and
$\mathbf{e_{\theta}}=\frac{1}{\epsilon r}\partial_{\theta}$ are unit basis
vectors.

To make the minimization of Frank free energy analytically tractable, our discussion is
limited to the torsion free tube whose core loop is a planar curve. Furthermore, we work
in the constraint that the radial component of the director field is zero
 and the nematic texture has axial
symmetry~\cite{kulic2004twist}. That is, $n_2(\alpha,r,\theta) = 0$, and $\mathbf{n}$ is
independent of $\theta$.  Therefore,
$\mathbf{n}=n_1(\alpha,r)\mathbf{e_{\alpha}}+n_3(\alpha,r)\mathbf{e_{\theta}}$.  Since
$\mathbf{n}$ is a unit vector, the simplest case is $\mathbf{n}=\mathbf{e_{\alpha}}$, i.e., all
the lines of the director are along the core loop of the tube.

Under these prescribed constraints, we derive for the following Euler-Lagrange
equation:
\begin{align}
\partial_k \frac{\partial
f}{\partial (\frac{\partial n_i}{\partial x_k})}+\frac{\partial_k
\sqrt{g}}{\sqrt{g}}\frac{\partial f}{\partial (\frac{\partial n_i}{\partial
x_k})}-\frac{\partial f}{\partial n_i}=-\lambda n_i,\nonumber
\end{align}
The condition of $n_3=0$ requires $\kappa'(\alpha)$ to be 0. In other words,
the core loop of the elliptic tube must be a circle to satisfy the Euler-Lagrange
equation. Such a circular torus solution turns out to be a minimum of the Frank free energy
by the following numerical analysis. 

Consider an elliptic core loop, after some calculation, we find that only the bending
term in Eq.(\ref{F}) is nonzero: 
\begin{equation} 
F=K_3 \pi\epsilon^2 \int_{0}^{2\pi}
\frac{\kappa(t)^2}{1+\sqrt{1-\epsilon^2 \kappa(t)^2}} (\sqrt{a^2
\textrm{sin}^2t+b^2 \textrm{cos}^2t}dt) \label{F_SI}
\end{equation} where $\kappa=ab/(a^2 \textrm{sin}^2t+b^2
\textrm{cos}^2t)^{\frac{3}{2}}$. $a$ and $b$ are the semi-major and semi-minor axes.
Numerical analysis of Eq.~(\ref{F_SI}) shows that $F$ monotonously decreases with $b$.
Since $b \leq a$, the optimal shape is therefore a circular torus. By inserting $b=a$ in
Eq.(\ref{F_SI}), we have $F_{\textrm{min}}/(K_3 \pi \epsilon^2) =
2\pi/(a+\sqrt{a^2-\epsilon^2})$.

\section{Acknowledgement}

This work was supported by NSFC Grants No. 16Z103010253, the SJTU startup fund
under Grant No. WF220441904, and the award of the Chinese Thousand Talents
Program for Distinguished Young Scholars under Grant No.16Z127060004 and
No. 17Z127060032.

\end{document}